# Phase Separating Microdomains in Highly Coarse-grained Lipids

S. Siva Nasarayya Chari

*Department of Physics, Faculty of Science, University of Allahabad, Prayagraj – 211002, Uttar Pradesh, India.*

Corresponding author: snchari@allduniv.ac.in

**Abstract.** A highly coarse-grained model for the lipid monolayer is formulated by considering a two-dimensional system of spherical particles with distinct types of S, U, and C, corresponding to the saturated, unsaturated, and cholesterol molecules. Their interactions are defined as modified Lennard-Jones type with distinct interaction strengths. The phase behavior of the system is studied for various cholesterol compositions, $x$ ranging from 0.1 to 0.6. From the simulation results, we observe that the S- and C-type particles segregate together to form ordered microdomains, with U-type particles at their outer leaflet; consistent with observations in literature. An order parameter $\Phi$ is defined to quantify the extent of phase separation among the constituents. The time averaged value of $\Phi$ is found increasing with the $x$ value. Radial distribution function calculations indicate the structural ordering within the phase separated domains.

## INTRODUCTION

The fundamental cellular processes like signal transduction and inter-cellular transport relate to the lateral organization of lipid molecules within the membrane layer. Initially the membrane monolayer was perceived to be a fluid-like mixture of its constituents [1]. Later in the twentieth century, it was understood that the lipid molecules in a monolayer would laterally organize themselves to form structurally ordered microdomains, known as the liquid-ordered ($l_0$) phase, that coexist with the unordered, fluid-like environment, the liquid-disordered ($l_d$) phase [2,3]. It is known that $l_0$ phase consists mainly of cholesterol and the sphingolipids, whereas the $l_d$ phase is rich in unsaturated lipids [4,5]. These ordered domains are technically termed as *lipid rafts* [6,7] which play a crucial role in cell signaling, and trafficking. Recent experimental observations reveal that the tendency of these domain formation increases with the cholesterol composition [8] until an optimal value. Here, we formulate a highly coarse-grained model for a lipid monolayer to study the domain formation within the monolayer at a given cholesterol composition. In this, a monolayer consists of saturated, unsaturated lipids and cholesterol molecules, corresponding to the S-, U- and C-type spherical particles respectively. More details of the model and results are discussed in the following sections.

## METHODS AND RESULTS

The S- and U-type particles are assumed to be of same size, $\sigma_S = \sigma_U = 1.0$, whereas the C-type is considered to differ by 1% without loss of generality, $\sigma_C = 1.01$. Interactions between the S-C types are of modified Lennard-Jones potential with its interaction strength proportional to the cholesterol composition [9], $\epsilon_{SC} = \epsilon_{CS} = x$,

$$V_{\alpha\beta}(r) = V_{LJ}^{\alpha\beta}(r) - V_{LJ}^{\alpha\beta}(r_c),$$
$$V_{LJ}^{\alpha\beta}(r) = 4x\left[\left(\frac{\sigma_{\alpha\beta}}{r}\right)^{12} - \left(\frac{\sigma_{\alpha\beta}}{r}\right)^{6}\right]. \quad (1)$$

In the above, $\alpha, \beta \in \{S, C\}$, and the cut-off value $r_c = 2.5\sigma_{\alpha\beta}$. Whereas the interactions involving the U-type are of Weeks-Chandler-Andersen type with $r_c = 2^{1/6}\,\sigma_{U\beta}$, where $\beta \in \{S, C, U\}$. The ternary mixture of S-, C- and U-type particles (total $N = 1000$) is initially taken in a two-dimensional simulation box at a desired density of $\rho^* = 0.3$. Independent simulations were performed for various compositions of C-type, $x \in \{0.1 \text{ to } 0.6\}$, in steps of 0.1. The S- and U-type are equally distributed in number, conserving the total, $n_S + n_U + n_C = N$. The system is then thermalized with soft potential and then equilibrated with the above specified interactions. The Berendsen thermostat at the reference temperature $T^* = 0.1$, and the coupling constant $\tau = 10\delta t$ is employed for sampling NVT ensemble configurations of the system. The equations of motion of the system were time-integrated up to 2 ns, with time step $\delta t = 2$ fs. Snapshots of the system at different C-type compositions are depicted in Fig. 1.

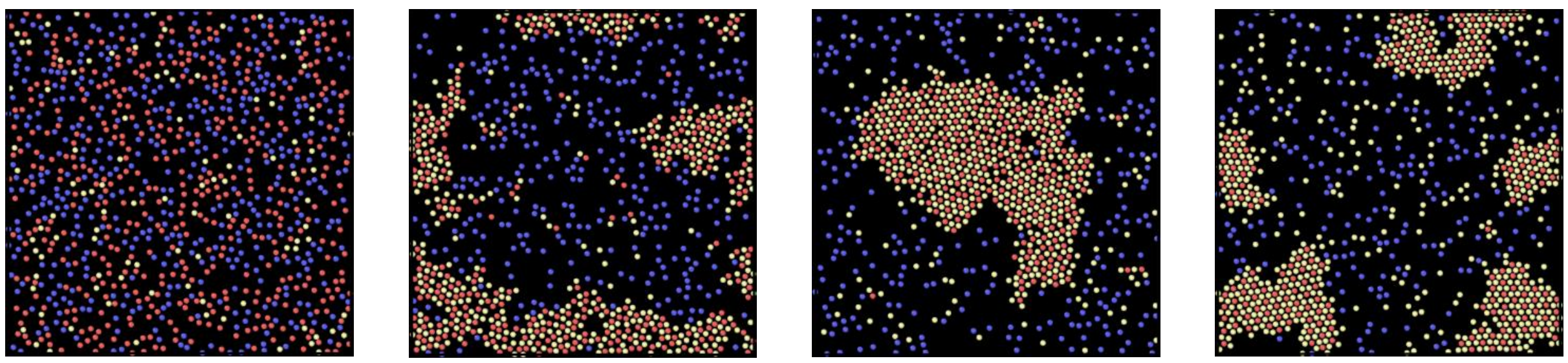

(a) $x = 0.1$ (b) $x = 0.4$ (c) $x = 0.5$ (d) $x = 0.6$

**FIGURE 1.** System snapshots after equilibration, at the specified C-type compositions. Color code is as follows: blue, red, and yellow corresponds to U-, S-, and C-type respectively.

From Fig. 1 we notice that the S- and C-type segregate together to form structurally ordered domains with U-type fluid-like medium at their outer leaflet. The phase separation order parameter Φ [10] is employed to quantify the extent of the domain formation. The entire simulation box is divided into number of cells where size of the cell is much greater than the size of the particle, $\Delta \gg \sigma_{ij}$. We used $Ncells = 16$ in this calculation.

$$\Phi(t) = \sum_{i}^{Ncells} \frac{|\, n_{S,C}{}^{i} - n_{U}{}^{i} \,|}{(n_S + n_U + n_C)^{i}} \tag{2}$$

In the above, $n_{S,C}^i$ and $n_U^i$ refer to the population of S, or C-type, and U-type particles in i-th cell respectively. We then sum over all the cells to get a measure for the phase separation in the entire system at any instant of time.

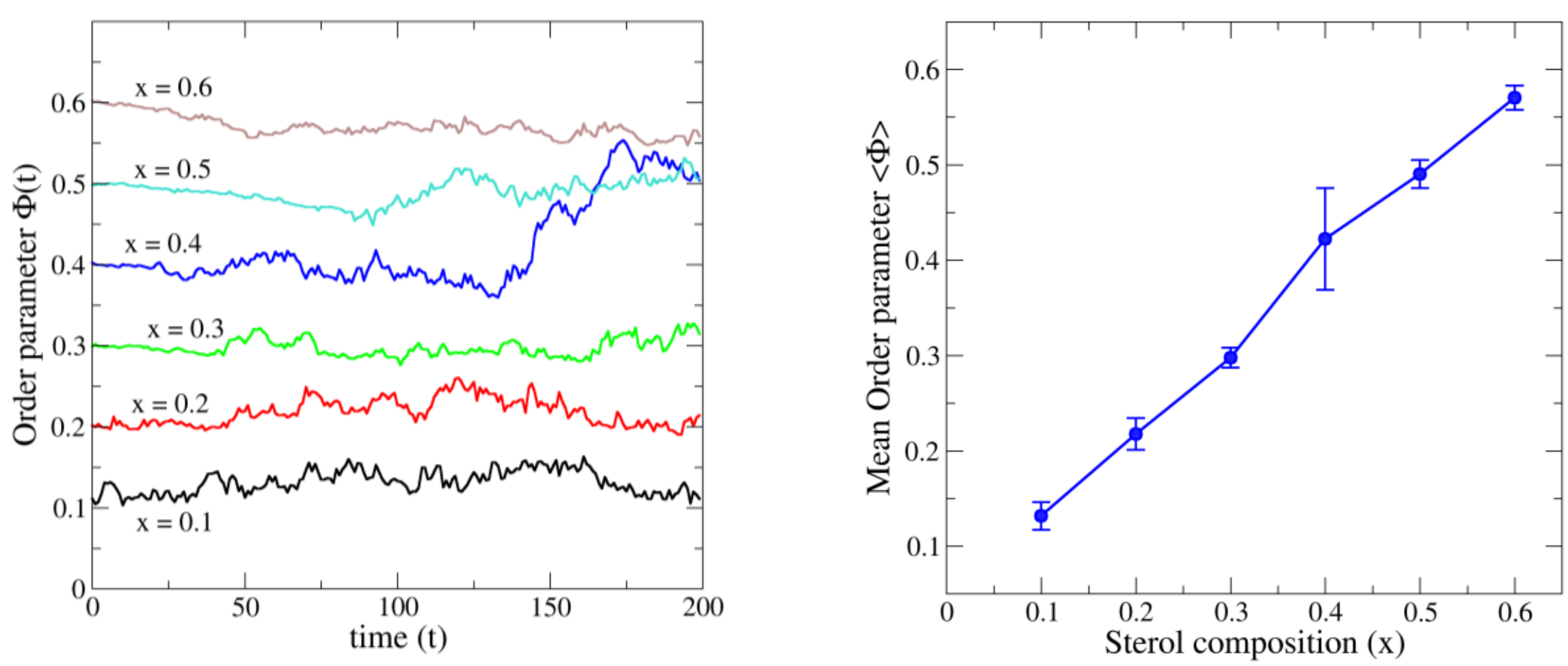

(a) (b)

**FIGURE 2.** (a). Phase separation order parameter for various compositions of C-type particles, and (b). the ensemble averaged order parameter at the specified values of $x$.

This is further averaged over all saved equilibrium configurations of the system to obtain the mean order parameter,

$$\langle \Phi \rangle = \frac{1}{\Omega} \sum_{j=1}^{\Omega} \Phi(j). \tag{3}$$

From Fig. 2(a), we notice the time variation of the phase separation order parameter for different $x$ values. Figure 2(b) shows that the mean order parameter increases linearly with the C-type composition. Radial distribution function $g_{sc}(r)$ is obtained between the S- and C-types, as shown in Fig. 3. This ensures that the phase separated domain is indeed structurally ordered, for $x = 0.6$.

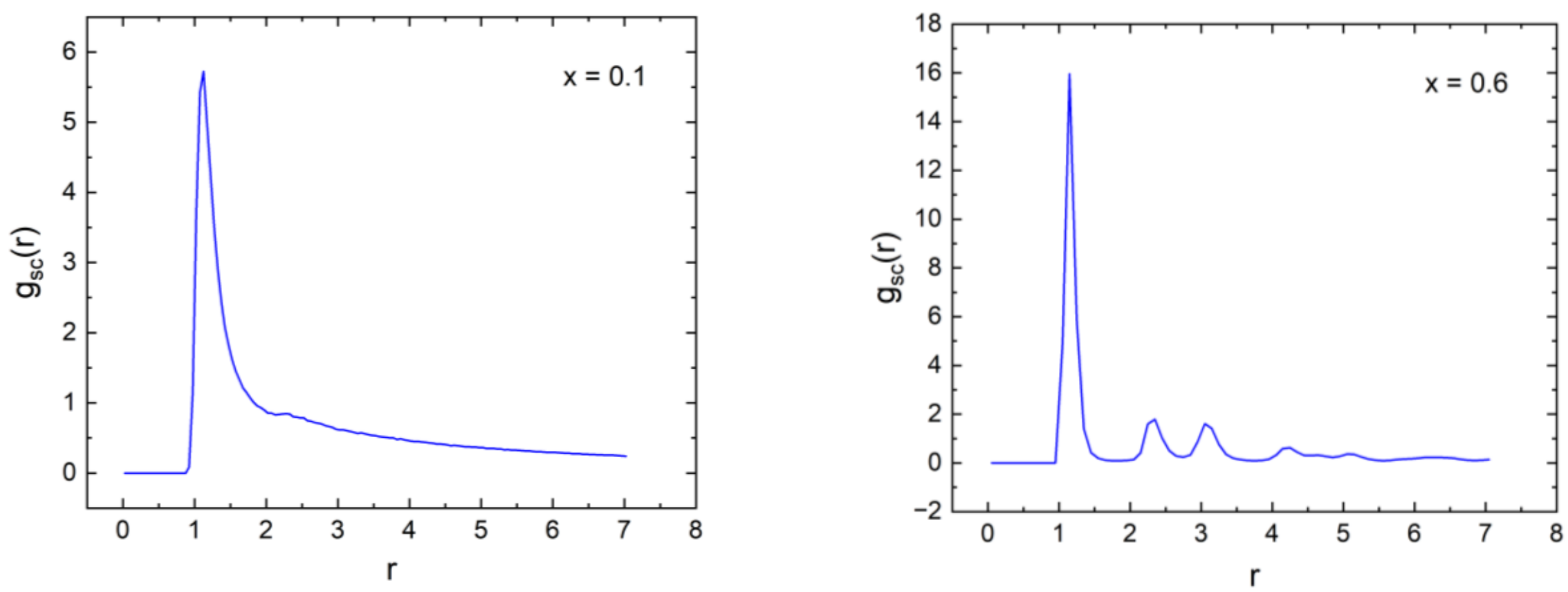

(a) (b)

**FIGURE 3.** Radial distribution function calculated between the S-C types at the compositions of $x = 0.1$ and 0.6.

To summarize, a highly coarse-grained model is formulated to study the dynamical segregation of the lipids in presence of cholesterol at various compositions of the latter. The interaction strength between S-C types is taken as a linear function of the cholesterol composition. This brings the model a mean field approach of the three-body type interactions, that are recommended for studying the ternary mixtures of saturated, unsaturated lipids, and cholesterol molecules. The extent of the phase separation between the S-, C- and U-type particles is quantified through the defined order parameter Φ. The mean order parameter shows a linear increment with the C-type composition throughout the range $x \in \{0.1 - 0.6\}$.